\documentclass[preprint,epsf,aps,floats]{revtex4}
\usepackage{amsmath,amssymb,bm,epsfig,psfrag}
\usepackage{mathtools}

\DeclarePairedDelimiter\abs{\lvert}{\rvert}%

\newcommand\+{\dagger}

\newcommand{\be}{\begin{equation}}
\newcommand{\ee}{\end{equation}}
\newcommand{\ber}{\begin{eqnarray}}
\newcommand{\eer}{\end{eqnarray}}

\newcommand\bra[1]{\langle #1 |}
\newcommand\ket[1]{|{#1}\rangle}

\def\Dsl{\,\raise.15ex \hbox{/}\mkern-12.8mu D}

\begin{document}

\title{Enhanced Multiple Exciton Generation in Amorphous Silicon Nanoparticles}
\author{Andrei~Kryjevski}
\affiliation{Department of Physics,~North Dakota State University,~Fargo, ND~58108,~USA}
\author{Dmitri Kilin}
\affiliation{Department of Chemistry,~University of South Dakota,~Vermillion,~SD~57069,~USA}
\begin{abstract}
Multiple exciton generation (MEG) in nanometer-sized hydrogen-passivated silicon nanowires (NWs), and quasi two-dimensional nanofilms 
strongly depends on the degree of the core structural disorder as shown by the many-body perturbation theory (MBPT) calculations 
based on the density functional theory (DFT) simulations. Working to the second order in the electron-photon coupling and in the 
screened Coulomb interaction we calculate quantum efficiency (QE), the average number of excitons created by a single absorbed photon, in the 
${\rm Si}_{29}{\rm H}_{36}$ quantum dots (QDs) with crystalline and amorphous core structures, 
simple cubic three-dimensional arrays constructed from these QDs, crystalline and amorphous NWs, and quasi two-dimensional 
silicon nanofilms, also both crystalline and amorphous. 
Efficient MEG with QE of 1.3 up to 1.8 at the photon energy of about $3E_g$, where $E_g$ is the 
electronic gap, is predicted in these nanoparticles except for the crystalline NW and crystalline film where $QE\simeq 1.$ 
MEG in the amorphous nanoparticles is enhanced by the electron localization due to structural disorder. Combined with the lower 
gaps, the nanometer-sized amorphous silicon NWs and films are predicted to have effective carrier multiplication within the solar spectrum range.

\end{abstract}

\date{\today}

\maketitle

\section{Introduction}

Quantum dots (QDs) are few nanometer-sized particles with size-tunable optical properties (see, {\it e.g.}, \cite{doi:10.1021/cr900289f,BurdaChemicalReviews2005}). 
In many applications 
individual QDs are aggregated, or assembled into ordered
arrays \cite{Weller_AFM-02,BawendiSuperlattices1995}.
In these structures, the strength of inter-QD electronic coupling, and, therefore, carrier transport characteristics 
depend on the array's composition, such as packing order, inter-QD distances 
and orientations, the possible surface shell structure, passivating  ligands, {\it etc.}
\cite{doi:10.1021/cr900289f,Talapin_Lee_Kovalenko_Shevchenko_2010,doi:10.1021/ja209062d}.
The limiting case of a densely packed one-dimensional (1D) array where individual QDs 
are merged corresponds to a nanowire (NW). Properties of NWs have also been actively investigated \cite{YunanXiaWIRES2003,PeidongYang_MaterialsToday_2006,ThelanderMT2006,PicrauxAPL2012,PicrauxAdvMat2011}.
%
To predict dependence of the optoelectronic properties of QD arrays and NWs on the chemical composition, 
surface and core structure and degree of spatial confinement
is a major challenge in the nanomaterial design for light and energy applications \cite{doi:10.1146/annurev.physchem.49.1.371,Talapin_Lee_Kovalenko_Shevchenko_2010}.  
 
Studying properties of QDs and NWs made of silicon have received a lot of attention since silicon is a material that is not only already in wide use, 
but also shows further promise in various applications \cite{ISI:000244069000054}.
In confined structures, such as QDs and NWs, the indirect gap nature of bulk crystalline silicon 
is modified which enhances their photophysical properties \cite{PhysRevLett.81.2803}. In particular, amorphous silicon NWs have been studied \cite{doi:10.1021/nl802886y}.

An important property of a nanoparticle is 
how effectively the energy of an absorbed photon will be converted into 
the energy of excited charge carriers.
Photon-to-electron energy conversion processes in nanoparticles have been under active investigation. This is due, in part, to the potential 
to increase the maximum theoretical efficiency of the nanomaterial-based solar cells
via 
carrier multiplication, 
or multiple exciton generation (MEG) process, where multiple excitons are created from one absorbed photon 
\cite{10.1063/1.1736034,ISI:000229120900009,AJ2002115}. Put another way, one strives to increase efficiency of the photon-to-electron energy 
conversion by diverting the excess photon energy into generation of extra charge carriers instead of losing it to atomic vibrations \cite{AJ2002115}.

A potent characteristic of the MEG process is the average number of excitons generated by an absorbed photon of a given energy. 
This quantity is called quantum efficiency (QE) (or, more appropriately, internal quantum efficiency), and can be measured in experiments \cite{Semonin16122011}. 
In this work we use QE to describe MEG.

In the solar photon energy range carrier multiplication has low efficiency in the bulk semiconductors 
\cite{5144200,5014421,10.1063/1.370658}.
But in nanoparticles, such as QDs and NWs, MEG efficiency is expected to be greatly increased due to enhancement of
electron Coulomb interactions by the spatial confinement \cite{doi:10.1146/annurev.physchem.52.1.193,AJ2002115,doi:10.1021/nl0502672,doi:10.1021/nl100177c}. 
However, the confinement will also increase the electron gap which will inhibit absorption at low photon energies. To mitigate this problem, 
one needs to explore possible mechanisms of lowering the gap while still retaining the beneficial effects of confinement. For instance, recently $Si$ and $Ge$ nanoparticles 
with the high-pressure bulk phase structure 
\cite{PhysRevLett.110.046804,C4TA01543F}, and $Si$ nanocrystals with reconstructed surfaces \cite{PhysRevB.87.155402}
have been proposed. In this work we propose to consider MEG in amorphous silicon nanoparticles, such as QDs, NWs and nanofilms, since they tend to have lower gaps compared the similar-sized crystalline counterparts.

In the course of investigation of MEG in semiconductor nanoparticles
drastically different opinions have been expressed regarding its mechanism and efficiency. See, {\it e.g.}, 
\cite{doi:10.1021/cr900289f,PhysRevLett.92.186601,doi:10.1021/nl0502672,KlimovMEGvirtualsingleexcitonNaturePhys,PhysRevB.78.125325,PhysRevB.76.081304}.
By now a consensus has emerged that MEG at energies higher than the $2 E_g$ threshold is, indeed, present in nanostructures, such as 
colloidal lead chalcogenide QDs \cite{doi:10.1021/nn800093v,doi:10.1021/nl100177c,doi:10.1021/nl0708617,doi:10.1021/jz200166y,doi:10.1021/nl803600v,Parkinson_Science_2010}.
In $Si$ QDs MEG with QE of 2.6 at the photon energy $3.4 E_g,~E_g$ is the energy gap, was reported \cite{doi:10.1021/nl071486l};
low threshold efficient MEG ($QE \simeq 1.5$ at $2.4 E_g$) was observed in silicon QDs (about 3.5 nm in diameter) dispersed in 
silicon dioxide \cite{trinh-2012}. 
Also, QE exceeding 100\% has been observed in $PbSe$ nanorods \cite{Sandberg:12}, and in the QD-based solar cells \cite{Semonin16122011}.
The need to describe both MEG and carrier energy relaxation due to phonon emission -- the two competing processes -- has been emphasized \cite{PhysRevB.88.155304,doi:10.1021/jz4004334}.

\begin{figure}[!t]
\center
\includegraphics[width=16cm]{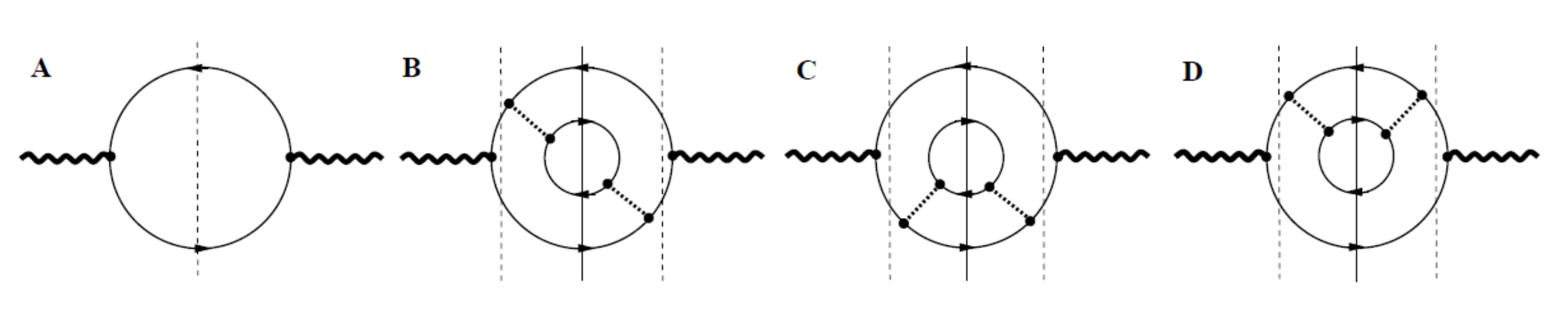} 
\caption{Photon self-energy ($\Sigma^{aa},~a={x},{y},{z}$) Feynman diagrams relevant for the single exciton ($R_1(\omega)$) and bi-exciton ($R_2(\omega)$) generation 
rate calculations.
Thin solid lines are the Kohn-Sham (KS) propagators, 
thick dashed lines are the screened Coulomb interactions, 
wavy lines are photons.
Dashed vertical line cuts correspond to single
exciton final physical states, solid line cuts correspond to bi-exciton states. 
Not shown 
are the vertex and KS self-energy corrections to the leading order 
process $A$ which are irrelevant under the approximations used in this work. 
(See Section \ref{sec:method}.) 
Processes $A,~B,~C,~D$ are described in the text.
}
\label{fig:graphs}
\end{figure}

Several carrier multiplication mechanisms have been considered.
In the impact ionization (I.I.) (or inverse Auger) process the 
electron or hole from an energetic exciton loses some of its energy 
via Coulomb interaction to create another 
exciton \cite{PhysRevLett.92.186601,doi:10.1021/nl049869w,PhysRevB.73.205423,doi:10.1021/nl0612401} (processes C, D in Fig. \ref{fig:graphs}, see also Fig. 1 of \cite{molphysKK}).
Another mechanism is the so-called direct photogeneration via virtual 
exciton states \cite{PhysRevB.76.125321,Rabani2010227,KlimovMEGvirtualsingleexcitonNaturePhys} (B in Fig. \ref{fig:graphs}, see also Fig. 1 of \cite{molphysKK}).
Direct Auger process, {\it i.e.}, the bi-exciton-to-exciton recombination, 
has, also, been studied \cite{doi:10.1021/nl0612401,PhysRevLett.106.207401,PhysRevB.87.155402}. 
It is understood that these carrier multiplication mechanisms are naturally included
in the many-body perturbation theory (MBPT) approach as particular contributions to the photon-to-exciton 
and the photon-to-bi-exciton processes computed to the second order in Coulomb interaction
\cite{Rabani2010227,PhysRevLett.106.207401,doi:10.1021/jz200166y} (see Fig. \ref{fig:graphs}, B, C, D, see also Fig. 1 of \cite{molphysKK}).
Note that MBPT approach naturally includes {bi-exciton-to-exciton} recombination, {\it i.e.}, the direct Auger process. 
For example, in Fig. \ref{fig:graphs}, D a photon with energy $\hbar\omega>2 E_g$ is absorbed and 
an electron-hole pair (exciton) is generated. The hole in this exciton then turns into a trion via Coulomb interaction. 
This trion and the original electron can form a physical bi-exciton state, or the trion can remain virtual eventually collapsing back 
into the hole state thus forming a single physical exciton.  

Going beyond perturbation theory, the coherent multiexciton production, where superposition of exciton and bi-exciton states is generated has
been considered \cite{doi:10.1021/nl0502672,doi:10.1021/nl062059v,PhysRevLett.105.137401}.

As already mentioned above, MEG is more effective in the smaller nanoparticles where on one hand
carrier interaction is confinement enhanced, and, on the other, where the role of the surface morphology, 
ligands, structural disorder is more important, as well. 
Currently, MEG in such systems can only be effectively described by the DFT combined with MBPT (MBPT$+$DFT).

To date, few MBPT studies of MEG in atomistic models of nanoparticles have been performed \cite{OssiciniNatPhot-2012,PhysRevLett.110.046804,PhysRevB.87.155402,molphysKK,C4TA01543F}. 
In \cite{OssiciniNatPhot-2012} I.I. MEG rates and bi-exciton lifetimes in the two $Si$ QD system of variable sizes were estimated;
 significant enhancement in carrier multiplication was observed as QD separation 
decreased. 
The space-separated quantum cutting scenario was proposed
where bi-excitons were directly generated on the adjacent 
QDs rather than first being produced on the same QD with 
subsequent inter-QD exciton transfer. Also, these delocalized 
bi-exciton states were found to have long recombination lifetimes.
I.I. rates in the isolated nanometer-sized $Si$ and $Ge$ nanoparticles
with core structures based on the high-pressure bulk phases were computed in \cite{PhysRevLett.110.046804,C4TA01543F}.
I.I. rates in $Si$ nanocrystals up to 2$~nm$ in size with reconstructed surfaces were studied in \cite{PhysRevB.87.155402}.
QE calculations have not been attempted in these studies. MBPT$+$DFT expressions for the 
photon-to-bi-exciton, $R_2$, and photon-to-exciton, $R_1$, rates needed for the QE calculation
have been derived in \cite{molphysKK}.
Also, in the same article the I.I. rates in the one-dimensional (1D) arrays of $Si_{29}H_{36}$ QDs 
have been calculated, and an enhancement was found as QD separation decreased \cite{molphysKK}.
{So far, exciton effects 
have not been included in any of the MBPT MEG calculations}. 

This work has two main objectives. 1.~We report calculations of the photon-to-bi-exciton, $R_2$, and photon-to-exciton, $R_1$, 
rates and the resulting QE in silicon nanoparticles using atomistic approach where DFT is combined with MBPT. To the best of our knowledge, 
this is the first attempt at such calculations. 
2.~The second goal is to to use this QE technique to study dependence of MEG in nanometer-sized hydrogen-passivated silicon QD arrays, NWs, and quasi two-dimensional 
nanofilms on the degree of core structural disorder, and on the QD separation 
which governs inter-QD interactions and formation of collective states. 

The paper is organized as follows. Section \ref{sec:method} contains brief overview of the methods
and description of the approximations employed in this work. Section III
contains description of the atomistic models studied in this work and of the DFT simulation details. Section \ref{sec:results} contains discussion of the results obtained. Conclusions and Outlook are presented in Section \ref{sec:conclusions}. 
\section{Theoretical Methods and Approximations}
\label{sec:method}
\subsection{Electron Hamiltonian in the KS basis}
\label{sec:H}
The annihilation operator of the $i^{th}$ Kohn-Sham (KS) state, ${\rm a}_{i\alpha},$ is related to the standard electron field 
$\psi_{\alpha}({\bf x})$ as
\ber
\psi_{\alpha}({\bf x})=\sum_i\phi_{i\alpha}({\bf x}){\rm a}_{i\alpha},
\label{psi_to_a}
\eer
where $\phi_{i\alpha}({\bf x})$ is the $i^{th}$ KS orbital, and $\alpha$ is the electron spin index \cite{FW,Mahan}. In this work we have only 
considered spin non-polarzed states with $\phi_{i\uparrow}=\phi_{i\downarrow}\equiv \phi_i$. 
The operators ${\rm a}$ and ${\rm a}^{\+}$ obey canonical anticommutation relations 
${\{}{\rm a}_{i\alpha},~{\rm a}_{j\beta}^{\+}{\}}=\delta_{ij}\delta_{\alpha\beta},~{\{}{\rm a}_{i\alpha},~{\rm a}_{j\beta}{\}}=0.$
In terms of ${\rm a}_{i\alpha}$ the electron Hamiltonian takes the form (see, {\it e.g.}, \cite{molphysKK})
\ber
{\rm H}=
\sum_{i\alpha}\epsilon_{i} {\rm a}_{i\alpha}^{\+}{\rm a}_{i\alpha}+
\frac{1}{8\pi}\int {\rm d}{\bf x} \left({\rm E}^2+{\rm B}^2\right)-\frac{1}{c}\int {\rm d}{\bf x}~{\bf j}\cdot{\bf A}+{\rm H}_{C}-{\rm H}_{V},
\label{H}
\eer
where $\epsilon_{i\uparrow}=\epsilon_{i\downarrow}\equiv \epsilon_i$ is the $i^{th}$ KS energy eigenvalue in the non-polarized case we consider here. The second term in Eq. \ref{H} is the standard photon Hamiltonian with ${\bf E}(t,{\bf x}),~{\bf B}(t,{\bf x})$ being electric and magnetic field operators, respectively. The third term in Eq. \ref{H} describes electron-photon coupling, where
\ber
{\bf j}({\bf x})=\left(\sum_{ij}\frac{e\hbar}{2 m}{\rm a}_{i\alpha}^{\+}
\left[\phi^*_i({\bf x})i{{\vec \nabla}}\phi_j({\bf x})\right]{\rm a}_{j\alpha}+h.c.\right)+{\cal O}({\bf A})
\label{Ae}
\eer
is the current operator, ${\bf A}(t,{\bf x})$ is the electromagnetic field vector potential operator, $e$ is the electron charge, ${\rm c}$ is the speed of light. The fourth term is
the (microscopic) Coulomb interaction operator 
\ber
{\rm H}_C=\frac12\sum_{ijkl~\alpha,\beta}{\rm V}_{ijkl}{\rm a}^{\dagger}_{i\alpha}{\rm a}^{\dagger}_{j\beta}{\rm a}_{k\beta}{\rm a}_{l\alpha},
~{\rm V}_{ijkl}=\int{\rm d}{\bf x}{\rm d}{\bf y}~\phi^{*}_i({\bf x})\phi^{*}_j({\bf y})\frac{e^2}{|{\bf x}-{\bf y}|}\phi_k({\bf y})
\phi_l({\bf x}).
\label{HC}
\eer
The last term is the compensating potential which prevents double-counting of electron interactions
\ber
{\rm H}_{V}=\sum_{ij}{\rm a}_{i\alpha}^{\+}\left(\int{\rm d}{\bf x}{\rm d}{\bf y}~\phi^*_i({\bf x}){V_{KS}({\bf x},{\bf y})}\phi_j({\bf y})\right){\rm a}_{j\alpha},
\label{HV}
\eer
where $V_{KS}({\bf x},{\bf y})$ is the, in general, non-local KS potential consisting of the Hartree and exchange-correlation terms (see, {\it e.g.}, \cite{RevModPhys.74.601,RevModPhys.80.3}). Coulomb gauge ${\bf \nabla}\cdot{\bf A}=0$ is used in this work.

Casting the electron Hamiltonian in the form (\ref{H}) makes it convenient to treat KS states as the effective Fermionic degrees of freedom (quasiparticles) with their interactions described by the last three terms of (\ref{H}). 
\subsection{Approximations}
\label{sec:approximations}
In this work we have used hybrid Heyd-Scuseria-Ernzerhof (HSE) exchange correlation functional in the DFT simulations \cite{vydrov:074106,heyd:219906}. Hybrid functionals have been somewhat successful in reproducing electronic gaps in various semiconductor nanostructures ({\it e.g.}, \cite{RevModPhys.80.3}). (See, however, \cite{PhysRevLett.107.216806}.) Here using HSE functional is assumed to substitute for calculating corrections to the KS energies $\epsilon_i$ using (costly) $GW$ method 
\cite{PhysRev.139.A796,PhysRevB.34.5390,0034-4885-61-3-002,RevModPhys.74.601}. So, here we use the simplest approximation where the single-particle energy levels 
and wave functions are approximated by $\epsilon_i$ and $\phi_i$ from the HSE DFT output. This implies that Fermion lines in Fig. \ref{fig:graphs} are assumed to be ``dressed'', {\it i.e.}, the self-energy corrections as well as the ${\rm H}_V$ term (\ref{HV}) are assumed to have been incorporated in the Fermion propagators \cite{RevModPhys.74.601}. While $GW$ (or, at least, $G_0W_0$) technique would have to be employed to improve accuracy of our calculations, it is unlikely to significantly alter our results and conclusions. 

In this work, electron-hole bound states (excitons) are approximated as uncorrelated KS particle-hole pairs. The bound state effects in the nanometer-sized silicon nanoparticles are important (see, {\it e.g.}, \cite{PhysRevB.68.085310,PhysRevLett.98.036807}). Incorporating them into this QE technique is in progress \footnote{A.~Kryjevski,~D.~Mihaylov,~D.~Kilin, unpublished}. 
However, bound state effects are not likely to change the conclusions of this work qualitatively.

Now the task is straightforward: use standard MBPT techniques ({\it e.g.}, \cite{AGD,FW}) to compute photon-to-bi-exciton and photon-to-exciton decay rates working to the second order in the electron-photon coupling and to the second order in the screened Coulomb interaction.

The effects of electron-phonon interactions are approximately taken into account 
by adding small imaginary parts to the KS energies
$\epsilon_i\rightarrow \epsilon_i - i \gamma_i$.
This is the source of non-zero line-widths in the expressions below. For simplicity, in this work all the line-width parameters will be set to $0.025~eV$ which corresponds to the room temperature scale. 

\subsection{Expressions for the rates $R_1,~R_2$ }

Let us start by quoting the KS orbital Fourier transformation conventions used in this work
\ber
&&\phi_i({\bf k}) = \frac{1}{\sqrt{V}}\int_{V} {\rm d}{\bf x}~\phi_i({\bf x}) {\rm e}^{-i{\bf k}\cdot{\bf x}},
~\phi_i({\bf x}) =\frac{1}{\sqrt{V}}\sum_{{\bf k}}\phi_i({\bf k}) {\rm e}^{i{\bf k}\cdot{\bf x}},\nonumber \\
&&{\bf k}
        =
          2\pi
          \left(
                \frac{n_x}{L_x},
                \frac{n_y}{L_y},
                \frac{n_z}{L_z}
          \right),~n_x, n_y, n_z=0,\pm 1, \pm 2,...
\label{phiKS}
\eer
with $V=L_x L_y L_z$ being the simulation cell volume. 

Generation of the exciton and bi-exciton states due to photon absorption can be viewed as a decay of the photon into 
the exciton and bi-exciton states, respectively. To the second order in the 
electron-photon coupling and in the screened Coulomb interaction the 
general expression for the 
photon
decay rate is
\ber
{\rm R}(\omega)=-\frac{4 \pi c^2 \hbar}{\omega V}{\rm Im}\left(\Sigma(\omega)\right),
\label{gamma_ph}
\eer 
where the polarization averaged optical photon self-energy, $\Sigma(\omega),$ 
is proportional to the time-ordered current-current correlation function
\ber
&& \Sigma(\omega)=\frac13\sum_a\Sigma^{aa}(\omega),~a={x},{y},{z}, \nonumber \\
&& i\Sigma^{ab}(\omega)=\frac{1}{\hbar^2 c^2}\int {\rm d} {\bf x} {\rm d} {\bf y} {\rm d}t~{\rm e}^{i \omega t}\bra{\Omega} {\rm T}j^{a}(t,{\bf x})j^b(0,{\bf y})
\ket{\Omega},
\label{sigma}
\eer
where ${\bf j}(t,{\bf x})$ is the current operator, Eq. (\ref{Ae}), ${\rm T}$ is the time ordering symbol, 
$\ket{\Omega}$ is the nanoparticle's ground state.
As mentioned above, the relevant Feynman diagrams 
are shown in Fig. \ref{fig:graphs}. Contributions to ${\rm Im}\left(\Sigma^{aa}(\omega)\right)$ from the dashed line cuts correspond to the single exciton 
final states and are relevant for ${\rm R}_1(\omega)$, while the solid line cuts corresponding to the bi-exciton physical states contribute to ${\rm R}_2(\omega).$ 
Then QE is given by \cite{Rabani2010227,PhysRevLett.106.207401}
\ber
QE(\omega)=\frac{R_1(\omega)+2 R_2(\omega)}{R_1(\omega)+R_2(\omega)}.
\label{QE} 
\eer

The leading order (LO) 
{photon}-to-{exciton} rate (Fig. \ref{fig:graphs}, A, dashed cut) is
\ber
{\rm R}^{LO}_1(\omega)=
\sum_{ij}\theta_j  \theta_{-i} 
\frac{8 \pi^2} {3 V\hbar \omega} 
|{\bf J}_{ji}|^2
\delta_{\gamma}(\omega-\omega_{ji}),
\label{R1LO}
\eer
where the summation notations are
\ber
\sum_{i}\theta_i=\sum_{i \geq LU},~\sum_{i}\theta_{-i}=\sum_{i \leq HO}
\label{thetai}
\eer 
with HO and LU defined as the highest occupied and lowest unoccupied KS levels,
and
\ber
&&{\bf J}_{ji}=e\sum_{{\bf k}}\phi^*_j({\bf k})\left({\frac{\hbar{\bf k}}{m}}\right)\phi_i({\bf k}),~\omega_{ji}=\frac{\epsilon_j-\epsilon_i}{\hbar},
\label{Jji}
\eer
where ${\bf k}$ is defined in Eq. (\ref{phiKS}),
and
\ber
\delta_{\gamma}(x)=\frac1\pi\frac{\gamma}{x^2+\gamma^2},
\label{delta_L}
\eer
the Lorentzian representation of the $\delta$-function. 
Note that (\ref{R1LO}) is proportional to the LO unpolarized optical photon absorption 
cross section, as prescribed by the optical theorem \cite{sakurai1994modern}. 
However, in the case of simulation cells of variable volume, the photon-to-exciton decay rate
${\rm R}_1(\omega)$ is a sensible measure of absorption.
For instance, in terms of ${\rm R}_1$ a sparse QD array and an isolated QD have the same absorption, as they should. Therefore, in this work we use ${\rm R}^{LO}_1$ 
from Eq. \ref{R1LO} as a measure of the absorption spectrum in a nanoparticle.

The LO contribution to $R_2(\omega)$ and the next-to-leading order (NLO) contribution to $R_1(\omega)$ from the I.I. process shown in Fig. \ref{fig:graphs}, C and D are
\ber
{\rm R}^{II}_2(\omega)&=&
\sum_{slkij} 
\frac{16 \pi^2} {3V\hbar^3 \omega}\abs*{{\bf J}_{sl}}^2 {\cal M}_{ksij}
\left(\theta_k {\theta_s}\theta_{-l} \theta_j \theta_{-i}
f^{p}_2+
\theta_{-k}\theta_{-s}\theta_l\theta_{-j} \theta_{i}f^{h}_2\right),\nonumber \\
f^{p}_{2}&=&
\left({\cal P}_{\gamma}({\omega-\omega_{sl}})\right)^2\delta_{\gamma}(\omega-\omega_{kl}-\omega_{ji}),
~f^{h}_{2}=
\left({\cal P}_{\gamma}({\omega-\omega_{ls}})\right)^2\delta_{\gamma}(\omega-\omega_{lk}-\omega_{ij}), 
\label{R2II}
\eer 
\vspace{-7ex}
\ber
{\rm R}^{II}_1(\omega)&=&
\sum_{slkij}
\frac{16 \pi^2} {3 V\hbar^3 \omega} \abs*{{\bf J}_{sl}}^2{\cal M}_{ksij}
\left(\theta_k {\theta_s}\theta_{-l} \theta_j \theta_{-i}
f^p_1+
\theta_{-k}\theta_{-s}\theta_l\theta_{-j} \theta_{i}
f^h_1\right),\nonumber \\
f^p_{1}&=&
2{\delta}_{\gamma}({\omega-\omega_{sl}}){\cal P}_{\gamma}(\omega-\omega_{kl}-\omega_{ji}){\cal P}_{\gamma}({\omega-\omega_{sl}}),\nonumber \\
~f^h_{1}&=&
2{\delta}_{\gamma}({\omega-\omega_{ls}}){\cal P}_{\gamma}(\omega-\omega_{lk}-\omega_{ij}){\cal P}_{\gamma}({\omega-\omega_{ls}}),
\label{R1II}
\eer
where
\ber
{\cal M}_{ksij}&=&
\abs*{\sum_{{\bf p}\neq 0}\frac{4 \pi e^2}{V}\frac{{{\rho}}_{ks}^{*}({\bf p}){{\rho}}_{ij}({\bf p})}{\left({p}^2-\Pi(0,{\bf p},-{\bf p})\right)}}^2,
\label{M}
\eer
and 
\ber
{\cal P}_{\gamma}(x)=\frac{x}{x^2+\gamma^2},~
{{\rho}}_{ji}({\bf p})=\sum_{{\bf k}}\phi_j^{*}({\bf k}-{\bf p})\phi_i({\bf k}),
\label{rhoij}
\eer
the transitional density,
and where
\ber
\Pi(\omega,{\bf k},{\bf p})=\frac{8 \pi e^2}{V\hbar}\sum_{ij}\rho_{ij}({\bf k})\rho_{ji}({\bf p})\left(\frac{\theta_{-j}\theta_{i}}{\omega-\omega_{ij}+i\gamma}-\frac{\theta_{j}\theta_{-i}}{\omega-\omega_{ij}-i\gamma}\right)
\label{Piwkp}
\eer
is the random phase approximation (RPA) polarization insertion (see, {\it e.g.}, \cite{FW}), and where the notations introduced in Eqs. {\ref{thetai},\ref{Jji},\ref{delta_L}} have been used.

Let us now describe simplified treatment of medium screening used in this work. We start with the standard RPA screened Coulomb potential
\ber
{\rm W}(0,{\bf k},{\bf p})=4\pi e^2\left[k^2\delta_{{\bf k},-{\bf p}}-\Pi(0,{\bf k},{\bf p})\right]^{-1},
\label{Wwkp}
\eer
where $\Pi(\omega,{\bf k},{\bf p})$ given by Eq. \ref{Piwkp}, in the static limit $\omega=0$ widely used for $Si$ nanostructures 
({\it e.g.}, \cite{PhysRevLett.90.127401,PhysRevB.68.085310,PhysRevB.79.245106}).
So, evaluating ${\rm W}(0,{\bf k},{\bf p})$  
requires matrix inversion. For $1-2$ nm-sized systems the matrix dimensionality can easily reach $10^{5}-10^6,$ and, more generally, the cost of this operation scales as as $N^3{\mathrm ln}N,~N$ is the number of atoms, which 
can limit applicability of the MBPT techniques \cite{Deslippe20121269,PhysRevB.79.245106}. 
A significant technical simplification can be achieved by retaining only the {\it diagonal} matrix elements in $\Pi(0,{\bf k},{\bf p}),$ {\it i.e.}, approximating 
$\Pi(0,{\bf k},{\bf p})\simeq\Pi(0,{\bf k},-{\bf k})\delta_{{\bf k},-{\bf p}}$ as implemented in Eq. \ref{M}. In position space this corresponds to 
$\Pi(0,{\bf x},{\bf x^{'}})\simeq\Pi(0,{\bf x}-{\bf x^{'}}),$
{\it i.e.},
to approximately treating the system as a uniform medium.

To determine quality of this approximation we have used our code to solve Bethe-Salpeter Equation (BSE) 
\cite{PhysRev.139.A796,PhysRevB.34.5390,0034-4885-61-3-002,PhysRevB.29.5718,PhysRevB.62.4927} for crystalline $Si_{29} H_{36},~Si_{35} H_{36}$ QDs using ``diagonal'' 
static interaction ${\rm W}(0,{\bf k},-{\bf k}).$ DFT simulations were done using B3LYP \cite{doi:10.1021/j100096a001} and HSE06 functionals, respectively. Then 
low-energy absorption including exciton effects was calculated, and the optical gap, $E_{opt},$ was determined. The results are shown in Table 1.
{\begin{table}
\raisebox{0.00001\totalheight}{\begin{tabular}{|c|c|c|}\hline
\textbf{Structure} & $cSi_{29} H_{36}~B3LYP$ & $cSi_{35} H_{36}~HSE06$ \\ \hline
\textbf{$E_g,~eV$}&5.0&4.4\\ \hline
\textbf{$E^{BSE}_g,~eV$}&3.1&2.6\\ \hline
\textbf{$E_{opt},~eV$}&4.5&4.4 \\ \hline
\end{tabular}}
\label{Pik-kresults}
\caption{$E_g$
is the HO-LU gap, $E^{BSE}_g$ - BSE minimal exciton energy, $E_{opt}$ - optical gap.} 
\end{table}}
Comparison with the existing high precision calculations for $Si_{29} H_{36},~Si_{35} H_{36}$ QDs has shown that optical gaps, $E_{opt}$, agree with the results of Garufalis {\it et~al.} \cite{GarufalisPRL2001} within  few \%. The minimal exciton state energy, $E^{BSE}_g,$ for $Si_{35} H_{36}$ agrees with with the results of Benedict {\it et~al.} \cite{PhysRevB.68.085310} within  few \% (no data for $Si_{29} H_{36}$). Our $E_{opt}$ are lower by about 10\% compared to the Quantum Monte Carlo (QMC) results of \cite{PhysRevLett.89.196803} which may be due to the lack of single particle energy corrections in our calculations.

This indicates that the simplified screened interaction ${\rm W}(0,{\bf k},-{\bf k})$ used in this work is a reasonable approximation for the $nm$-sized hydrogen-terminated silicon nanoparticles. 
Using the more expensive full interaction ${\rm W}(0,{\bf k},{\bf p}),$ or ${\rm W}(\omega,{\bf k},{\bf p})$, will be needed to improve the accuracy of this QE technique but it will not significantly change the results or alter the conclusions of this work. 

\begin{figure}[!t]
\center
\includegraphics[width=16cm]{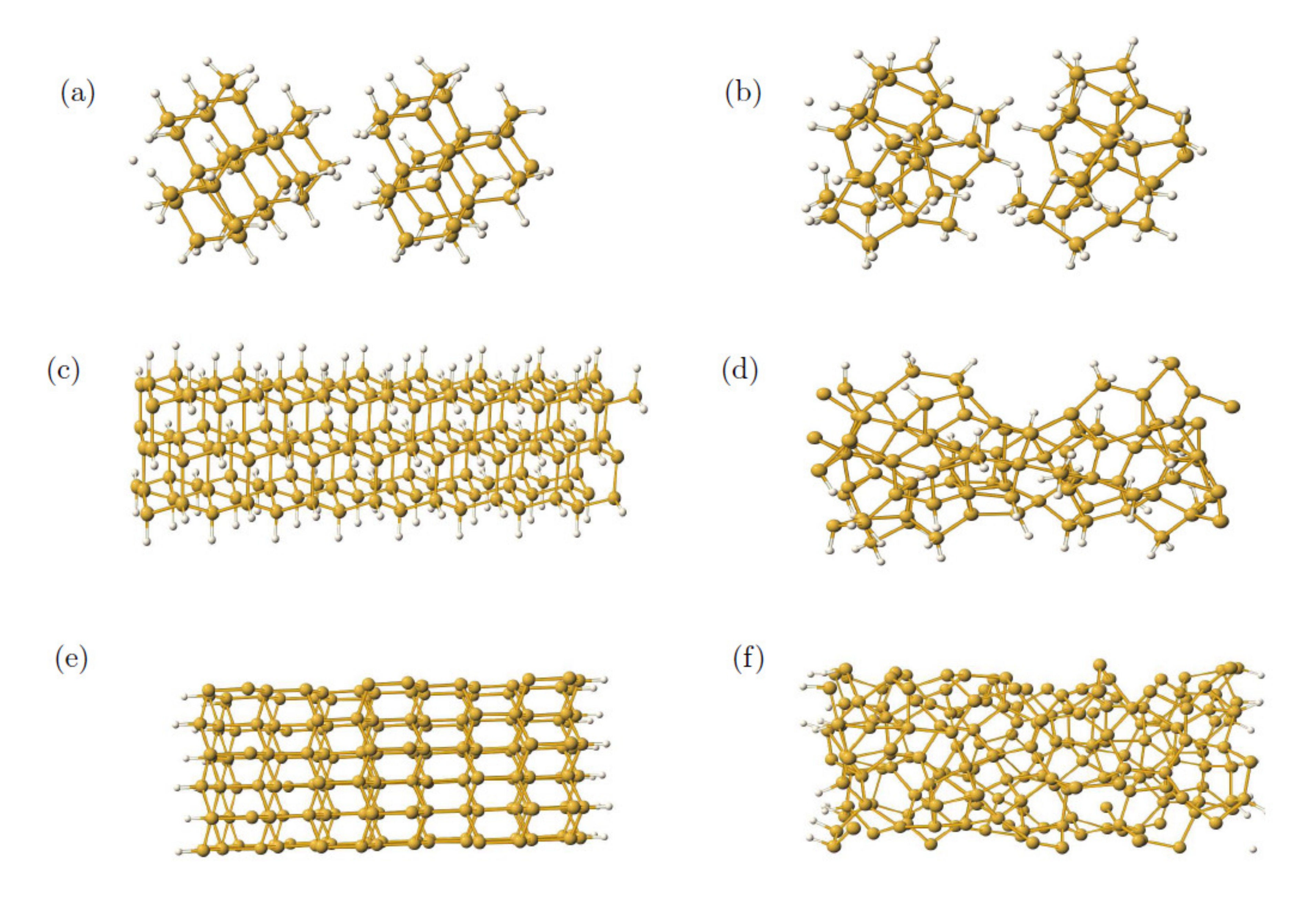} 
\caption{Unit cells of the DFT optimized atomistic models of nanoparticles. 
The core (yellow) atoms are silicons, the surface passivating (white) atoms are hydrogens. 
Shown in (a) and (b) are the two QD unit cells of the crystalline and amorphous $Si_{29} H_{36}$ QD arrays, respectively. 
In (c) is the crystalline $\langle112\rangle$ NW (${\rm Si}_{144}{\rm H}_{96}$, which is two unit cells).
In (d) is one of the amorphous NWs ($\sim 1~nm$ cross-section size, ${\rm Si}_{96}{\rm H}_{44}$). In (e) and (f) are the
crystalline and amorphous films, respectively ($\sim 2~nm$ thick, ${\rm Si}_{192}{\rm H}_{24}$). 
}
\label{fig:optimized_structures}
\end{figure}

Now let us finish quoting the rate expressions. Contribution to 
$R_2$ from the diagram shown in Fig. \ref{fig:graphs},~B (the vertex correction) is given by
\ber
&&{\rm R}^{VC}_2(\omega)=
\sum_{ijklab}
\frac{16 \pi^2} {3V\hbar^3 \omega} {\bf J}_{ij}\cdot{\bf J}_{kl}
\left(\sum_{{\bf p},{\bf q}\neq 0}
\frac{4 \pi e^2{{\rho}}_{ba}^{*}({\bf p}){{\rho}}_{jk}({\bf p})}{V\left({p}^2-\Pi(0,{\bf p},-{\bf p})\right)}
\frac{4 \pi e^2{{\rho}}_{il}^{*}({\bf q}){{\rho}}_{ba}({\bf q})}{V\left({q}^2-\Pi(0,{\bf q},-{\bf q})\right)}\right)
\times
\nonumber \\&&
\times\theta_j \theta _{-l}\theta_{-i} \theta_k\theta_a \theta_{-b}\delta_{\gamma}(\omega-\omega_{ab}-\omega_{jl})
P_{\gamma}(\omega -\omega _{{ji}})P_{\gamma}(\omega -\omega _{{kl}})+c.c.
\label{R2VC}
\eer
The corresponding ${\rm R}^{VC}_1(\omega)$ is given by a similar expression which is not shown here. This is because we have calculated ${\rm R}^{VC}_2(\omega),~{\rm R}^{VC}_1(\omega)$ for several nanoparticles from Fig. \ref{fig:optimized_structures} and found their magnitudes to be only few \% of the corresponding I.I. contributions to the rates, ${\rm R}^{II}_2(\omega)$ and ${\rm R}^{II}_1(\omega).$ 
This agrees with the earlier findings that MEG is dominated by the I.I. process \cite {PhysRevLett.106.207401}. So, from now on we will approximate
\ber
{\rm R}_1(\omega)={\rm R}^{LO}_1(\omega)+{\rm R}^{II}_1(\omega),~{\rm R}_2(\omega)={\rm R}^{II}_2(\omega),
\label{R12}
\eer
where ${\rm R}^{LO}_1(\omega),~{\rm R}^{II}_1(\omega),~{\rm R}^{II}_2(\omega)$ are given by Eqs. \ref{R1LO},\ref{R1II},\ref{R2II}, respectively.
As mentioned above, ${\rm R}^{II}_1(\omega)$ receives contributions from the dashed line cuts in Fig. \ref{fig:graphs} C, D that correspond to the 
single exciton final states.
These include the {bi-exciton-to-exciton} state recombination, {\it i.e.}, the direct Auger process. 

Note that in Eqs. \ref{R1II},\ref{R2II},\ref{R2VC} the zero momentum mode
is excluded from the 
momentum sums which is due to charge neutrality \cite{FW}.    

Pauli exclusion principle in Eqs. \ref{R2II},\ref{R2VC} has been implemented by excluding contributions from the physical bi-exciton states with two particles or 
holes with the same spin occupying the same KS state. 

We emphasize that the rates $R_1,~R_2$ are comprehensive characteristics of the photon-to-exciton and the photon-to-bi-exciton conversion processes, respectively. They 
naturally include both the initial photon absorption process
and the subsequent exciton-to-bi-exciton conversion via medium-modified Coulomb interaction.  
{\section{Computational Details}}
\label{sec:compdetail}

The optimized geometries and the electronic structures of the silicon nanostructures   
have been obtained using the {\it{ab-initio}} total-energy and
molecular-dynamics program VASP (Vienna ab-initio simulation program) 
with the hybrid Heyd-Scuseria-Ernzerhof (HSE) exchange correlation functional \cite{vydrov:074106,heyd:219906}
along with the 
projector augmented-wave (PAW) pseudopotentials \cite{PhysRevB.50.17953,PhysRevB.59.1758}.

VASP uses plane wave basis, and the momentum cutoff is implemented
by the condition
\ber
\frac{\hbar^2{k}^2}{2 m}\leq {\cal E}_{max},
\label{Ecutoff}
\eer
where ${\bf k}$ is a finite-volume discretized wave vector (Eq. \ref{phiKS}),
$m$ is the electron mass. 
In our simulations ${\cal E}_{max}=312.5~eV$ 
has been used.
Conjugated gradient method for ionic relaxation available in VASP has been used. The geometries were relaxed until 
residual forces on the ions were no greater than $0.03~eV/\AA.$ 
The energy cut-offs regulated by the number of KS orbitals included into simulations were chosen so that 
$\epsilon_{i_{max}}-\epsilon_{HO}\simeq\epsilon_{LU}-\epsilon_{i_{min}}\geq 4 E_g,$ where $i_{max},~i_{min}$ are the highest and the lowest KS labels included in simulations.
The amorphous structures have been prepared by simulated annealing \cite{doi:10.1021/jp2055798,jrse}.
\begin{figure}[!t]
\center
\includegraphics[width=16cm]{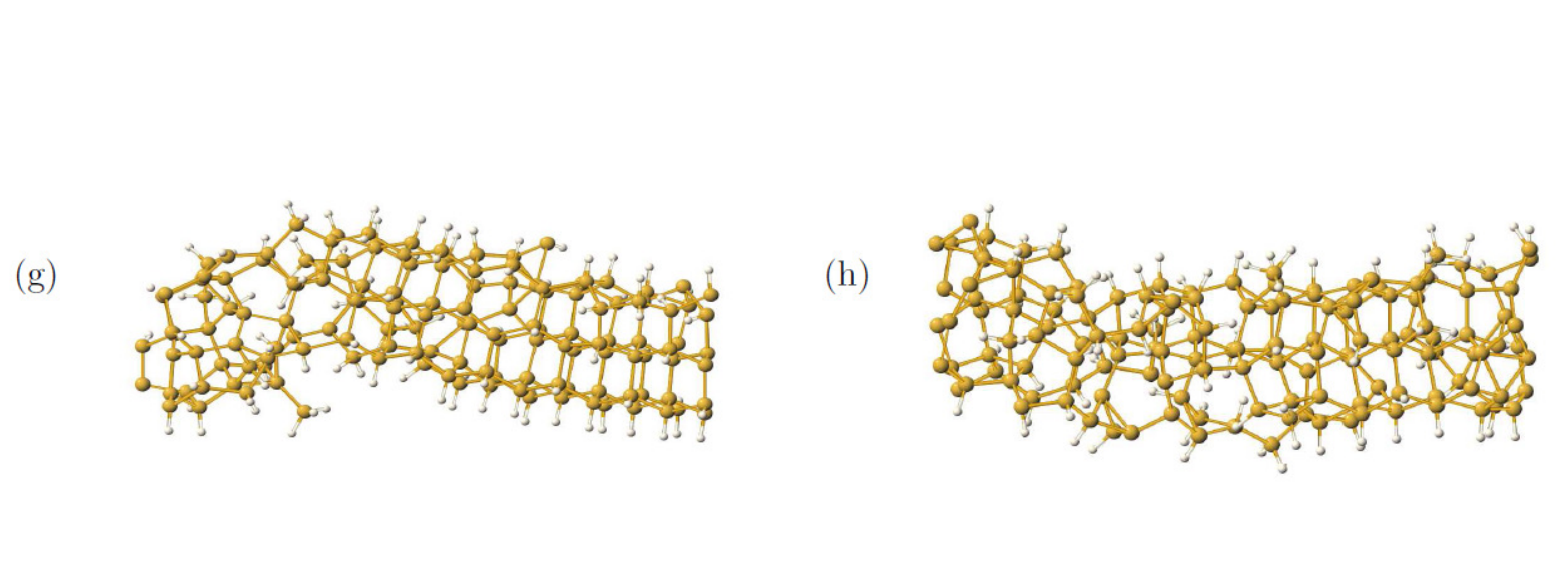} 
\caption{DFT relaxed units cells of the atomistic models of partially amorphous ((g), ${\rm Si}_{144}{\rm H}_{98}$) and amorphous ((h), ${\rm Si}_{144}{\rm H}_{88}$) NWs obtained by the simulated annealing of the crystalline NW (Fig. \ref{fig:optimized_structures}, (c)). 
}
\label{fig:pa_a_NWs}
\end{figure}

Periodic silicon nanostructures considered in this study have been represented 
by the atomistic models 
placed in various finite volume simulation boxes with periodic boundary conditions. 
For instance, both crystalline and amorphous isolated QDs which are about $1~nm$ in size have been simulated in cubic 
boxes with $L_x = L_y = L_z \simeq 2~nm.$ This has ensured inter-QD separation by about $1~nm$ of 
vacuum thus excluding spurious interactions between their periodic images.
3D simple cubic QD arrays (\ref{fig:optimized_structures}, (a) and (b)) have been prepared by placing two QDs 
replicated in the $x$ direction into a box with dimensions $L_x = 2.5~nm,~L_y = L_z \simeq 1.25~nm.$
Upon relaxation this has resulted in the $\sim1.25~nm$ distance between QD geometric centers, which corresponds to the minimum energy 
simple cubic array configuration (as predicted with the HSE06 functional). 
For the NWs the length of the box has been chosen to coincide with the unit cell length, while in the 
other two directions the NWs have been kept separated by 
about $1~nm$ of vacuum. For the crystalline NW studied here (Fig. \ref{fig:optimized_structures}, (c)) the periodicity direction is $\langle 112 \rangle$ direction 
of Si bulk (two unit cells, $Si_{144}H_{96},$ have been simulated). It has approximately $1~nm$ cross-section size.

Five atomistic models of NWs with disordered or partially disordered core structure have been prepared and simulated in this work. Four of them have approximately $1~nm$ cross-section size, while the fifth one has turned out to have approximately elliptical cross-section with the dimensions of about 1.4~nm and 2.2~nm. (See Section \ref{sec:results} for more information.)

The quasi 2D films have been simulated by placing their unit cells into periodic boxes with the two dimensions close to the dimensions of the unit cell
(in-out of the page and up-down directions in Fig. \ref{fig:optimized_structures}, (e), (f)). In the third direction (left-right in Fig. \ref{fig:optimized_structures}, (e), (f))) the hydrogen passivated surfaces have been kept separated by about $1~nm$ of vacuum \cite{doi:10.1021/jp808908x}. For the crystalline film the $(111)$ surface is exposed.

In the preparation of disordered structures we have been careful to passivate the nanoparticle surfaces and to perform geometry relaxation to avoid 
artificial reduction of the gap by the unpaired surface electrons, and by other artifacts.   

The resulting gaps are shown in Table 2.
The gap, $E_g,$ which in our approximation is $E_g=\epsilon_{LU}-\epsilon_{HO},$ diminishes with the decrease of spatial confinement, {\it i.e.}, as one proceeds from the isolated QDs to arrays, then to NWs, and to the films. This is as expected and is due to the increase in the electron delocalization. In the amorphous nanostructures disorder results in the formation of multiple defects which introduces additional states in the band gap region, leading to the gap decrease in the nanoparticles with the core disorder compared to their crystalline counterparts.  
\begin{table}
\raisebox{0.00001\totalheight}{\begin{tabular}{|c|c|c|c|c|c|c|c|c|c|c|}\hline
\textbf{Structure} & cQD $Si_{29} H_{36}$ & aQD $Si_{29} H_{36}$ & (a) & (b) & (c) & (d), (h) ave. & (e) & (f) & (g) & aNW $Si_{200}H_{88}$\\ \hline
\textbf{$E_g,~eV$}& 4.51 & 3.64 & 4.45 & 3.67 & 2.09 & 0.90 & 1.46 & 0.55 & 1.59 & 0.61\\ \hline
\end{tabular}}
\label{table:Eg}
\vspace{-1.1ex}
\caption{Gaps for the structures shown in Figs. \ref{fig:optimized_structures}, \ref{fig:pa_a_NWs}, and for the amorphous NW $Si_{200}H_{88}.$}
\end{table}

Simulations of periodic structures have been done for ${\bf K}=0,$ where ${\bf K}$ is the lattice 
wave vector, {\it i.e.}, at the $\Gamma$ point. We have studied effects of the Brillouin zone sampling by including more $K$ points in the simulation of
the two QD unit cell of the 1D crystalline QD array (Fig. \ref{fig:optimized_structures}, (a)), and for the crystalline NW (Fig. \ref{fig:optimized_structures}, (c)) 
with two unit cells included in the simulation. We have found small (less than 10\%) variation in the single particle energies 
over the Brillouin zone \cite{jrse}. Based on this we have concluded that our $\Gamma$ point approximation is reasonable. So, in this work a KS orbital is specified by just an integer.
Full ${\bf K}$ dependence would have to be included to improve accuracy of our calculations. 

DFT often fails to correctly describe dispersion interactions \cite{Kristyán1994175}.  
To check this issue, we have conducted geometry optimization of the 1D QD arrays (cells from Fig. \ref{fig:optimized_structures}, (a), (b)), both 
crystalline and amorphous, using functional with van der Waals corrections, 
such as DFT-D2 method of Grimme \cite{ISI:000241477200003} included in VASP software.
Introduction  of dispersion corrections to DFT calculations has resulted in insignificant 
changes in the relaxed geometries, as well as in the density of states and absorption spectra \cite{jrse}.
 
In this work all DFT simulations have been done in a vacuum. Our calculations can serve 
as the simplest models of arrays of QDs, NWs dispersed in a low permittivity dielectric host 
material, such as SiO$_2$ (bulk dielectric constant is $\epsilon=3.9,$ the gap, $E_g\simeq 9~eV$).
In a more sophisticated approach the $SiO_2$/Si interface effects should be taken into account.
But investigation of these effects would require a large scale atomistic level study (see, {\it e.g.}, \cite{PhysRevLett.107.206805,doi:10.1080/00268976.2013.836606}) without significantly changing 
the results and conclusions of this work. 
\section{Results and Discussion}
\label{sec:results}
\begin{figure}[!t]
\center
\includegraphics[width=16cm]{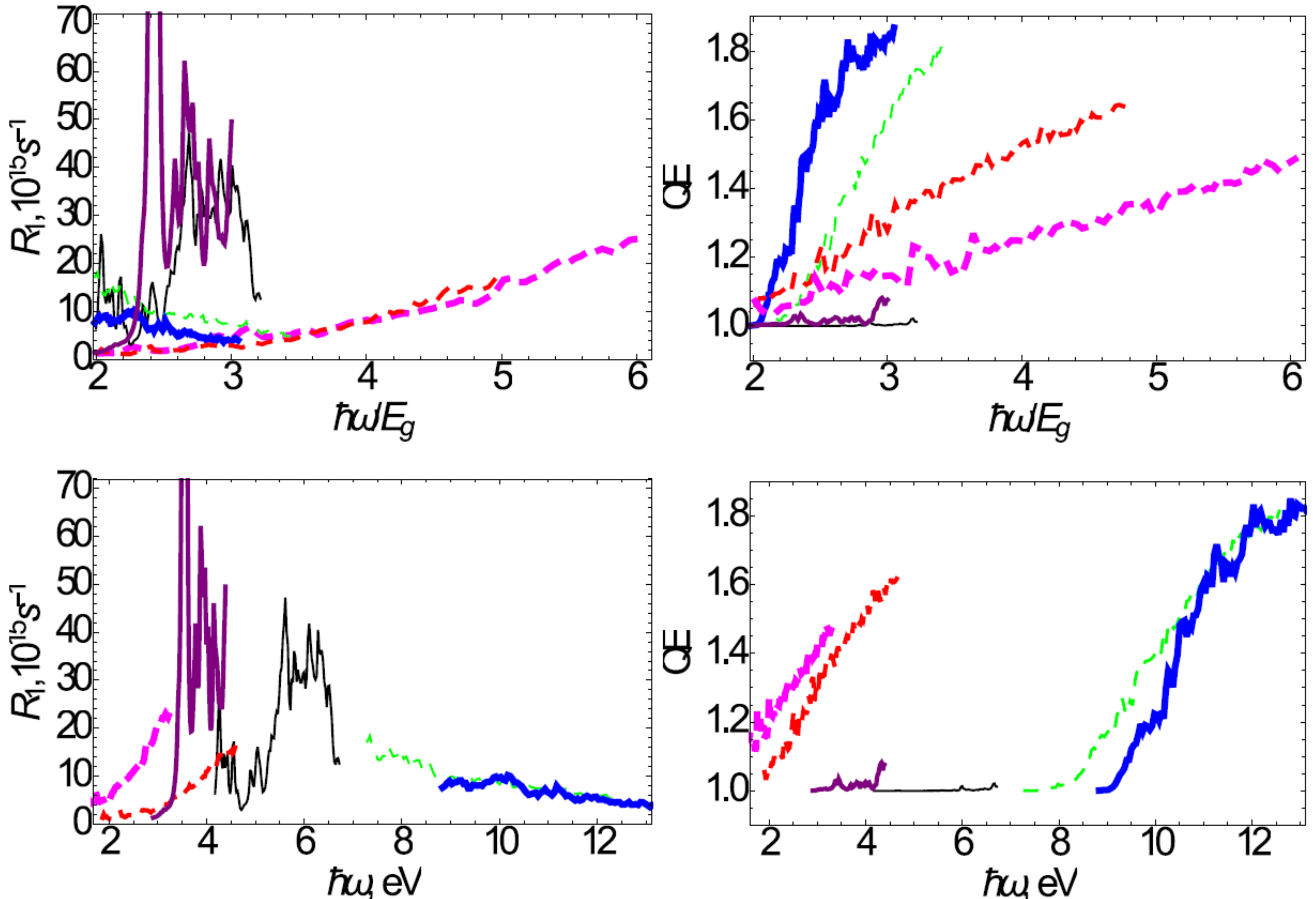} 
\caption{Absorption spectra and QEs for the six nanoparticles shown in Fig. \ref{fig:optimized_structures}. Thick solid (blue) line depicts 3D crystalline QD array (a), thin dashed (green) line - 3D amorphous QD array (b), thin (black) solid line -- crystalline NW (c), intermediate dashed (red) line -- amorphous NW, averaged, intermediate solid (purple) line -- crystalline film (e), thick dashed (magenta) line -- amorphous film (f). 
}
\label{fig:R1_QE_all}
\end{figure}

Shown in Fig. \ref{fig:R1_QE_all} on the left are the absorption rates 
for the structures shown in Fig. \ref{fig:optimized_structures}. When plotted as functions of $\hbar\omega/E_g$
the crystalline NW and film have the strongest absorption. However, as a function of the absolute energy the strongest absorption in the low-energy region $\hbar\omega\leq 3.5~eV$ is displayed by the amorphous film and NWs, which is due to the lower $E_g$ (see Table 2). Shown in Fig. \ref{fig:R1_QE_all} on the right are the QE results for the structures from Fig. \ref{fig:optimized_structures}. When plotted as a function of $\hbar\omega/E_g$ it is the QDs and QD assemblies, both crystalline and amorphous, that have the highest QE of about 1.8 at $\hbar\omega/E_g\simeq 3$.
As noted in Section III,
we have performed simulations on the isolated QDs
and on the simple cubic 3D arrays. The QEs are quite similar in both cases, so only the 3D array results are shown, for brevity. The QDs are followed by the amorphous NWs with average $QE\simeq1.6$ at $\hbar\omega/E_g\simeq 4.8$ and by the amorphous nanofilm with $QE \simeq 1.5$ at $\hbar\omega/E_g\simeq 6$. Our calculations indicate that MEG in the crystalline NW and in the crystalline nanofilm is very weak with $QE\simeq1$. As a function of the absolute energy the strongest MEG at low energy is exhibited by the the amorphous film ($QE\simeq1.5$ at $\hbar\omega=3.3~eV$) and by the 1nm cross-section NWs (average $QE\simeq1.6$ at $\hbar\omega \simeq 4.5~eV$) due, again, to the lower $E_g$ (see Table 2). 

In all cases including Auger recombination, {\it i.e.}, using ${\rm R}_1={\rm R}^{LO}_1+{\rm R}^{II}_1$ as opposed to ${\rm R}_1={\rm R}^{LO}_1,$ has reduced QE by not more 
than 5 \%.  

So, our results predict effective MEG 
in all the structures considered in this work except for the crystalline NW and the quasi 
2D crystalline film. This is not surprising and is similar to the low MEG efficiency in 
the periodic bulk semiconductors: delocalized plane wave-like states have electrostatic 
interactions that are too weak for an effective MEG.

QDs used in the existing QE measurements ($d=9.5~nm$ in \cite{doi:10.1021/nl071486l}, $d=3.5~nm$ 
in \cite{trinh-2012}) are much larger than the 1 nm ones we used in our simulations. 
Since MEG is very size-sensitive direct comparison is not possible at this time. Our 
crystalline QD array prediction of $QE=1.5$ at photon energy 2.4$E_g$ appears to agree 
with the measurements of \cite{trinh-2012}. We believe that this is a fluke. 
\subsection{MEG in NWs as a function of structural disorder}
\label{sec:MEGNWorder}
\begin{figure}[!t]
\center
\includegraphics[width=16cm]{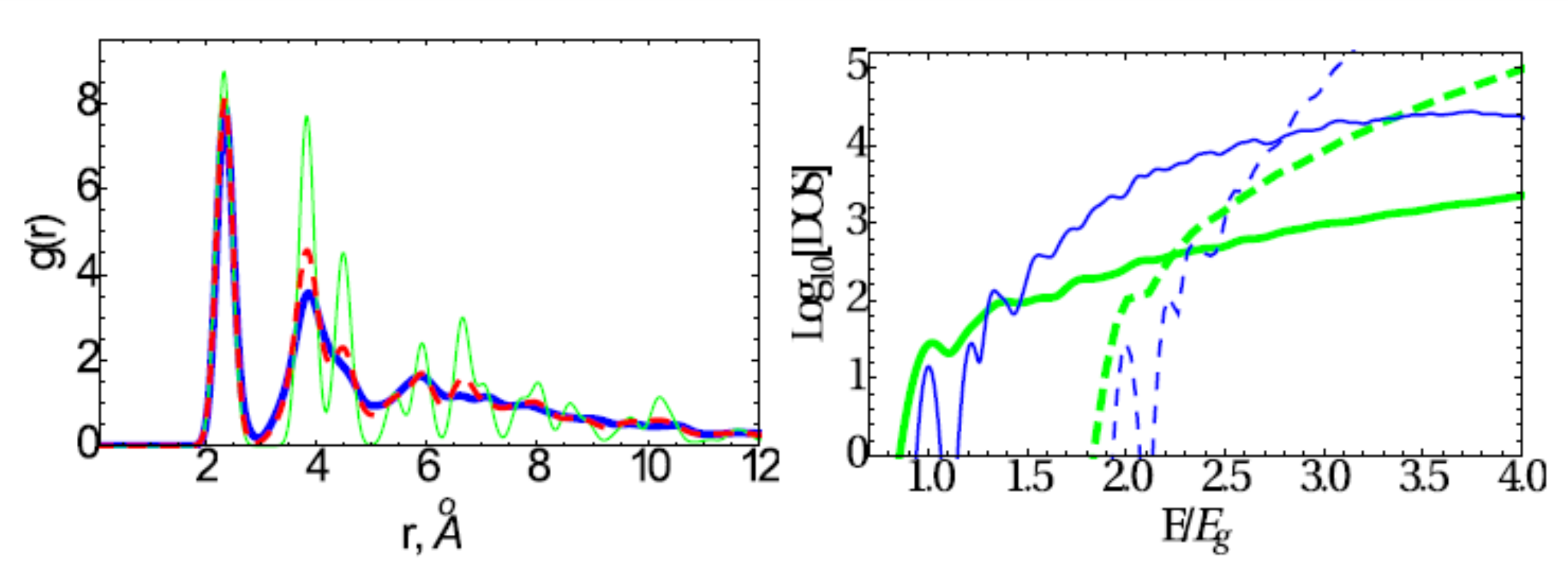} 
\caption{{\bf On the left}: Pair correlation functions for the crystalline NW (Fig. \ref{fig:optimized_structures}, (c)) depicted by the thin solid (green) line), partially amorphous 
(Fig. \ref{fig:pa_a_NWs}, (g)) -- dashed (red) line, 
and amorphous NW (Fig. \ref{fig:pa_a_NWs}, (h)) -- thick solid (blue) line. {\bf On the right}: Exciton and bi-exciton densities of states (DOS) for the $Si_{96}H_{44}$ amorphous NW, Fig. 
\ref{fig:optimized_structures}, (d) (thick solid line depicts exciton DOS, thick dashed line - bi-exciton DOS), and for the $Si_{144}H_{96}$ crystalline NW, 
Fig. \ref{fig:optimized_structures}, (c) (thin solid line - exciton DOS, thin dashed line - bi-exciton DOS). Gaussian broadening parameter $0.08~eV$ has been used.
}
\label{fig:PCF_c_pa_a_excDOS_cNW_aNW}
\end{figure}
\begin{figure}[!t]
\center
\includegraphics[width=16cm]{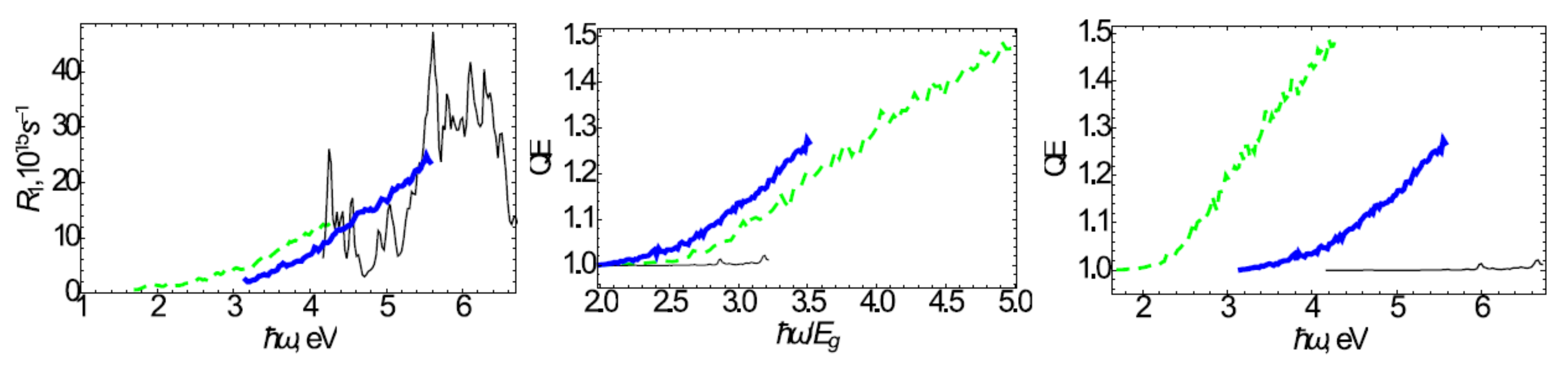} 
\caption{Absorption and QE of the crystalline 
(Fig. \ref{fig:optimized_structures}, (c), ${\rm Si}_{144}{\rm H}_{96},$ thin (black) solid line), partially disordered (Fig. \ref{fig:pa_a_NWs}, (g), ${\rm Si}_{144}{\rm H}_{98},$ thick (blue) solid line) and fully disordered (Fig. \ref{fig:pa_a_NWs}, (h), ${\rm Si}_{144}{\rm H}_{88},$ dashed (green) line) NWs.
}
\label{fig:R1_QE_c_pa_a_NW}
\end{figure}

To elucidate the dependence of MEG on the degree of core structural disorder in the NWs we have 
performed simulated annealing on the crystalline NW (Fig. \ref{fig:optimized_structures}, (c)) to make
one NW with a partially disordered structure (Fig. \ref{fig:pa_a_NWs}, (g)) and another with a fully disordered NW (Fig. \ref{fig:pa_a_NWs}, (h)). 
To quantify the disorder we have computed pair correlation functions 
\ber
{\rm g}(r)=\frac{1}{4 \pi r^2 N \rho}\sum_{i=1}^{N}\sum_{k \neq i}^{N}\delta(r-|{\bf x}_i-{\bf x}_k|),
\label{PCF}
\eer
where $N$ is the number of core atoms in the unit cell ($N_{Si}=144$ in our case), $\rho$ is the number density, for these three atomistic models. 
The results are shown in Fig. \ref{fig:PCF_c_pa_a_excDOS_cNW_aNW}, left panel. The nearest neighbor peaks at $r \sim 0.24$ nm are virtually identical for both crystalline and disordered structures. This is as expected since this peak characterizes the $Si-Si$ bond length in a solid silicon material. However, at greater separations there are noticeable differences as one proceeds from the crystalline (thin solid (green) curve) to the partially disordered (dashed (red) curve) and the fully disordered (thick solid (blue) curve) NW.
The peaks 
become broader and, overall, the curves exhibit fewer features. All this indicates evolution from the crystalline to amorphous structure in the three nanoparticles considered. 

So, shown in Fig. \ref{fig:R1_QE_c_pa_a_NW} are the absorption spectra and QEs for the three structures. The amorphous and partially amorphous NWs exhibit strong absorption at 
low energies which is due to the lower $E_g$ (see Table 2). As a function of the absolute energy the strongest MEG at low energy is exhibited by the the fully amorphous NW (Fig. \ref{fig:pa_a_NWs}, (h)) with $QE\simeq1.5$ at $\hbar\omega=4.3~eV$). This is due to the gap hierarchy of the three structures (see Table 2). 
\begin{figure}[!t]
\center
\includegraphics[width=16cm]{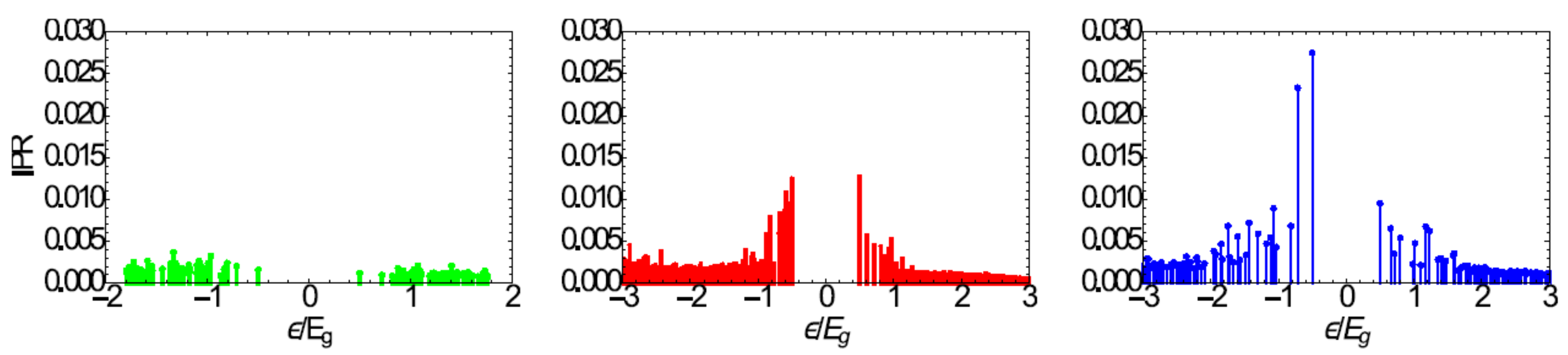} 
\caption{IPRs for the states near Fermi level for 1.~the crystalline 
(Fig. \ref{fig:optimized_structures}, (c), ${\rm Si}_{144}{\rm H}_{96}$, on the left, green lines), 2.~partially amorphous (Fig. \ref{fig:pa_a_NWs}, (g), ${\rm Si}_{144}{\rm H}_{98}$, in the middle, red lines) and 3.~amorphous (Fig. \ref{fig:pa_a_NWs}, (h), ${\rm Si}_{144}{\rm H}_{88}$, on the right, blue lines) NWs. Zero energy is set at mid-gap.
}
\label{fig:IPR_c_pa_a}
\end{figure}

To obtain an insight into the trend suggested by these results we have computed inverse participation ratios (IPRs) defined as
\ber
IPR_i=\frac{\int{\rm d}{\bf x}|\phi_i({\bf x})|^4}{\left(\int{\rm d}{\bf x}|\phi_i({\bf x})|^2\right)^2},
\label{IPR}
\eer
for the KS orbitals near Fermi level for the three NWs. IPR is a measure of localization of a single-particle state \cite{iprWegner}. The resulting IPRs are shown in Fig. 
\ref{fig:PCF_c_pa_a_excDOS_cNW_aNW}, left panel. As one proceeds from the crystalline to partially disordered, and then to amorphous NW the electron states become much more localized which enhances electron Coulomb interactions. 
In addition, shown in Fig. \ref{fig:PCF_c_pa_a_excDOS_cNW_aNW}, right panel are the exciton and bi-exciton densities of states (DOS) for the crystalline NW (Fig. 
\ref{fig:optimized_structures}, (c)) and for one of the amorphous NWs (Fig. \ref{fig:optimized_structures}, (d)). The crystalline NW suffers from a paucity of states
in the crucial regions near the energy thresholds $E=E_g$ and $E=2E_g$ which is an exhibition of the band structure. This explains drastically different MEG 
efficiencies in the crystalline and amorphous NWs and nanofilms. 
\subsection{Dependence of MEG in amorphous NWs on spatial confinement}
\label{sec:MEGNWconfine}
\begin{figure}[!t]
\center
\includegraphics[width=16cm]{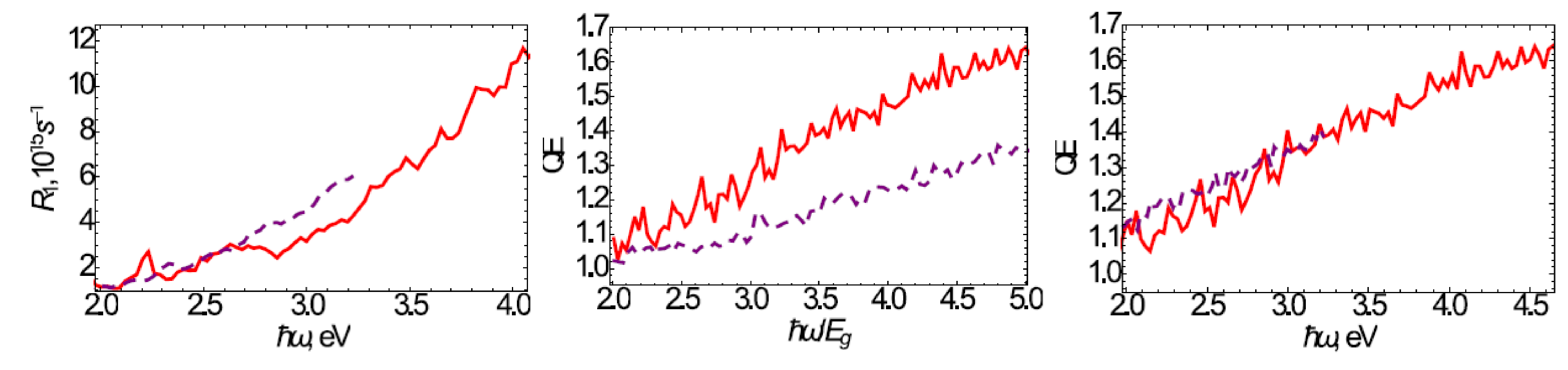} 
\caption{Absorption and QE of an amorphous ${\rm Si}_{100}{\rm H}_{50}$ NW (1~nm cross-section size, solid (red) line) and of a bigger ${\rm Si}_{200}{\rm H}_{88}$ NW (elliptical cross-section with 1.4~nm and 2.2~nm dimensions, dashed (purple) line).
}
\label{fig:R1_QE_Si100_Si200}
\end{figure}
Finally, to study how MEG in the amorphous NWs depends on the spatial confinement we have simulated a bigger disordered NW. 
This structure has been prepared by merging two copies of the $Si_{100}$ amorphous NW unit cell
($\sim1~nm$ cross-section size) with the subsequent hydrogen passivation and DFT geometry relaxation. The cross section of the resulting NW with $Si_{200}H_{88}$ unit cell 
has turned out to be approximately elliptical with the dimensions of about 1.4~nm and 2.2~nm. We have checked that the structures 
of $Si_{100}H_{50}$ and $Si_{200}H_{88}$ are very similar. Shown in Fig. \ref{fig:R1_QE_Si100_Si200} are the absorption rates and QEs for the two NWs. Both are predicted to have appreciable absorption at low energy. 
As a function of $\hbar\omega/E_g$ the smaller NW has higher QE, as expected. However, the $QE$s as functions of $\hbar\omega$ are similar in the two cases due to lower $E_g$ in the bigger NW (see Table 2).
\section{Conclusions and Outlook}
\label{sec:conclusions}

Working to the second order in the electron-photon coupling and in the screened Coulomb interaction 
we have developed a DFT$+$MBPT technique for calculating QE. This method has been used to calculate 
QE in several nanometer-sized hydrogen-passivated silicon nanoparticles in different regimes of spatial confinement, 
such as isolated silicon QDs, QD arrays, NWs, and quasi 2D films with both crystalline and amorphous core structures.

Our results have indicated that MEG in these systems is dominated by the I.I. mechanism (processes C, D in Fig. \ref{fig:graphs}) 
which is agreement with previous work \cite{PhysRevLett.106.207401}.
Our results predict that efficient MEG is, in general, present in the $Si$ nanoparticles considered here.
As one proceeds from the isolated QDs and QD arrays to NWs and films the efficiency of MEG crucially 
depends on the core structural disorder. Both crystalline and amorphous QDs and QD arrays exhibit strong MEG (Fig. \ref{fig:R1_QE_all}). 
But in the NWs and nanofilms, which can be viewed as limiting cases of dense 1D and 2D QD arrays, respectively, 
MEG is virtually absent in the crystalline structures but is strong in the amorphous ones (Figs. \ref{fig:R1_QE_all},\ref{fig:R1_QE_c_pa_a_NW}). 
Combined with the lower electronic gaps in the amorphous NWs and nanofilms we predict efficient MEG in these nm-sized 
nanoparticles already in the solar spectrum range (Fig. \ref{fig:R1_QE_all}). Inclusion of the exciton effects into calculations will 
further red-shift the $QE(\hbar\omega)$ curves. Given that the electron transport is typically stronger in the NWs and films compared to QD arrays,
nanomaterials based on the amorphous $Si$ NWs and/or nanofilms could be suitable for the energy conversion applications.

We realize that for some of our nanoparticles, especially the QDs, the photon energies required for the effective MEG 
are well outside of the optical spectrum range. But one expects that the qualitative trends found here will persist for larger nanoparticles. 

Accuracy of our methods can be improved in several ways. The single particle energy corrections can be included using standard $GW$ scheme. 
Then they can be easily incorporated in the rate expressions quoted above. 
As already mentioned above, accurate description of MEG in the 1-2 nm silicon nanoparticles requires incorporating exciton effects \cite{PhysRevLett.98.036807}. This will be 
one immediate extension of this work.
Inclusion of the dynamics of the electron-hole bound states in the rates ${\rm R}_1,~{\rm R}_2$ will be achieved using standard MBPT techniques (see, {\it e.g.}, \cite{Beane:2000fx}). 

Also, in the I.I. process the typical energy in the screened Coulomb potential exceeds the gap and is not negligible. The role of the dynamical screening effects should  be investigated. 

On a more general note, it is understood that conclusions about MEG efficiency in a nanoparticle can only be made by comparison with the phonon relaxation efficiency \cite{doi:10.1021/jz4004334}. So, in addition to the rates ${\rm R}_1,~{\rm R}_2$ 
one needs to compute exciton relaxation times due to phonon emission, and, then, the rates of decay of a photon into 
low-energy excitons. At the atomistic DFT$+$MBPT level this task can be achieved by employing the finite temperature/real time technique of MBPT \cite{Landau10,PhysRevB.83.165306}, 
or the reduced density matrix method \cite{InerbaevJPCC2013,SchmitzJPCL2013}.

We stress that while the theoretical methods of this work can be improved, the main conclusion of the drastic difference of the MEG efficiency between the amorphous and crystalline silicon NWs and nanofilms will persist. 

%
%
\section{Acknowledgments}
A. K. acknowledges financial support from the US Department of Energy via grant DE-FG52-08NA28921, and ND EPSCOR NSF Fund EPS-0814442, and
use of computational resources of the Center for Computationally Assisted Science and Technology (CCAST) at North 
Dakota State University.
D. K. acknowledges the South Dakota Governor's Office of Economic Development and NSF award EPS0903804 for financial support,  
and  DOE, BES-Chemical Sciences and NERSC No. DE-AC02-05CH11231, allocation Award 86185 for providing computational resources. 
A. K. and D.K. acknowledge financial support for method development from the NSF grant CHE-1413614.
\bibliography{dftNSF2013}
\end{document}